\documentclass[12pt,a4paper,reqno]{article}

\usepackage{amsmath,latexsym,amssymb,amsthm}
\usepackage{graphicx}

\usepackage{amsmath}\usepackage{amssymb}
\usepackage{amsthm}
\newcommand{\linespace}{\vspace{\baselineskip}}

\newcommand{\upline}{\vspace{-\baselineskip}}

\newcommand{\be}{\begin{equation}}
\newcommand{\ee}{\end{equation}}
\renewcommand{\(}{\left(}
\renewcommand{\)}{\right)}
\newcommand{\pr}{^\prime}

\renewcommand{\c}{\mathbf{c}}

\newcommand{\x}{\mathbf{x}}

\newcommand{\B}{\mathbf{B}}

\newcommand{\operone}{\mathbf{1}}

\newcommand{\C}{\mathbb{C}}
\renewcommand{\H}{\mathcal{H}}

\newcommand{\1}{\mathbf{1}}
\newcommand{\ox}{\otimes}
\newcommand{\ol}{\overline}
\newcommand{\<}{\langle}
\renewcommand{\>}{\rangle}
\newcommand{\half}{\tfrac{1}{2}}

\newcommand{\quarter}{\tfrac{1}{4}}
\newcommand{\eighth}{\tfrac{1}{8}}

\newcommand{\impl}{\Longrightarrow}

\newcommand{\tr}{\operatorname{tr}}

\theoremstyle{plain} 
\newtheorem{theorem}{Theorem}[section]

\theoremstyle{definition}

\theoremstyle{remark}

\numberwithin{equation}{section}
\newcommand{\thmref}[1]{Theorem~\ref{#1}}

\begin{document}

\title{\bf{Compatibility of subsystem states}}

\author{Paul Butterley,$^1$ Anthony Sudbery$^2$ and Jason Szulc$^3$\\[10pt] \small Department of Mathematics,
University of York, \\[-2pt] \small Heslington, York, England YO10 5DD\\
\small $^1$ pb504@york.ac.uk \quad $^2$as2@york.ac.uk \quad $^3$js115@york.ac.uk}

\date{22 April 2005\\------\\\normalsize \textbf{In memoriam Asher 
Peres}}

\maketitle

\begin{abstract}
\item

We examine the possible states of subsystems of a system of bits or
qubits. In the classical case (bits), this means the possible marginal
distributions of a probability distribution on a finite number of binary
variables; we give necessary and sufficient conditions for a set of
probability distributions on all proper subsets of the variables to be
the marginals of a single distribution on the full set. In the quantum
case (qubits), we consider mixed states of subsets of a set of qubits;
in the case of three qubits, we find \emph{quantum Bell inequalities} --- necessary conditions 
for a set of two-qubit states to be the reduced states of a mixed state of three qubits. We conjecture that these conditions are also sufficient.

\end{abstract}

\section{Introduction}

What can we believe about some parts of a system without contradicting what
we believe about other parts? If the system is described by a set of
numbers, and our beliefs are the probabilities that these numbers take
given values, then a part of the system is described by a subset of the
numbers and our beliefs about it will be given by marginal probabilities
derived from the probability distribution of the full set of numbers.
The marginal distributions of different parts are constrained by the
fact that they all come from a single set of probabilities on the full
system. Bell's inequalities are an example of such constraints. The
conclusion of the EPR argument is that a single quantum system like an
electron has a set of numbers giving the results of all possible
measurements, even though these cannot be measured simultaneously.
Wigner \cite{Wigner:ineq} presented Bell's theorem by considering the
probabilities for subsets of electron observables which could be
measured simultaneously (either directly, or by measuring the electron's
partner in a singlet state), and showing that these subset probabilities,
if they derived from a single probability distribution on the
full set, would be constrained by inequalities which were not satisfied
by the predictions of quantum mechanics.

Other forms of Bell inequalities can also be
understood in this way, as compatibility conditions on the marginal distributions of 
subsets. Asher Peres  \cite{Peres:allBell} has considered this problem
in complete generality, bringing out its formidable computational
complexity. In this paper we solve the special case in which one
is given joint probability distributions for all proper subsets of a set
of binary variables, finding necessary and sufficient conditions for
these distributions to be the marginals of a single distribution on the 
full set. 

The motivation for this study is to investigate our initial question for
quantum systems. In this case our knowledge of the 
system is
represented by a mixed state, or density matrix, and our knowledge of a
part of the system is given by the reduced state, obtained by tracing
the full density matrix over the rest of the system. What are the
constraints on these reduced states? Our answer to the classical problem
yields a possible answer to the quantum question, as the conditions on
marginal probability distributions have immediate analogues for quantum
states of a finite set of qubits. They can be
translated into conditions on the density matrices of
proper subsets of the qubits, which we prove, in the case of a system of three qubits, to be necessary
for the density matrices to be the reductions of a (mixed) state of the
full set of qubits. We conjecture that these \emph{quantum Bell-Wigner
inequalities} are also sufficient conditions. For more than three qubits the corresponding conditions are not even necessary; this gives rise to new separability criteria (the \emph{generalised reduction criteria}) \cite{Bill:reduction}.

A still more general problem in classical probability, which was
introduced by George Boole \cite{Boole:Laws}, is to ask when a set of
real numbers $p_{ijk\cdots}$ can be simultaneous probabilities $P(E_i\,\& \,E_j\,\&
\,E_k\,\&\,\ldots)$ for some events $E_i$. This problem has been investigated by
Pitowsky \cite{Pitowsky:book, Pitowsky:range}, who has
shown \cite{Pitowsky:polytopes} that the problem of deciding whether the
relevant conditions are satisfied is NP-complete. The relation to the
problem considered here (and by Peres) is that we assume that the full
sample space is a Cartesian product of finite sets and that our events
$E_i$ are slices of this product. 

Work on this problem appears to have concentrated on  
a (discrete or continuous) infinity of real-valued variables, i.e.\ a
stochastic process, in which case there are no conditions other than
the obvious ones (see \eqref{obvious} below); the Kolmogorov-Daniell
theorem \cite{Kolmogorov} asserts essentially that if these are satisfied 
for all finite subsets of the variables, then there is a stochastic
process of which they are the finite-time marginals. The focus then is
on the range of possible processes having these marginals. This problem
for bipartite quantum states has been studied by
Parthasarathy \cite{Parthasarathy} and Rudolph \cite{Rudolph:marginal}.

The situation in the quantum problem for pure states is, in a sense, 
inverse to this. It is not at all easy to construct an overall pure 
state which has given marginals: there are other conditions to 
be satisfied in addition to the
obvious ones \cite{polygon, Atsushi:3qutrit, Bravyi, HanZhangGuo}, and there is 
usually only one state with these marginals (this is the generic situation
if one is given the reduced states of subsets containing more than half of the total
number of qubits  \cite{NickNoah:parts, Diosi:reconstruct}). This can be
interpreted \cite{NoahSanduBill:power} as meaning that irreducible
$n$-way correlation is exceptional in pure $n$-qubit states.

However, it is not surprising that the quantum pure-state problem should
be different from the general classical problem, since the classical pure-state
problem is also very different. Classically, a pure probability
distribution consists of certainty; its marginals are also pure, the
only conditions to be satisfied by them are the obvious compatibility
conditions \eqref{obvious}, and the marginals of singleton subsets
uniquely determine the overall distribution. The quantum analogue of the
non-trivial classical problem is to ask when a set of subsystem states
is compatible with a mixed overall state. For identical particles, this
problem has been much studied \cite{Coleman:book}, but the case of 
distinguishable particles has only recently received attention. One
approach to it is outlined in \cite{NickNoah:parts}: in this paper we
suggest another line of attack.

The paper is organised as follows. In section 2 we consider the classical problem and present necessary and sufficient conditions for compatibility of probability distributions on proper subsets of a finite set of binary variables. In section 3 we describe the quantum problem, prove necessary conditions for compatibility of reduced states of two-qubit subsystems of a system of three qubits, and show that the corresponding conditions are not necessarily satisfied for a system of more than three qubits. In an appendix we review other work on the quantum marginal problem. 

\newpage

\section{Classical marginals}

The general classical problem is as follows. Let
$S = \{X_1,\ldots,X_n\}$ be a set of random variables, with $X_i$ taking
values in a finite set $V_i$. Let $A,B,\ldots$ be a set of subsets of
$\{1,\ldots,n\}$, and let $S_A, S_B,\ldots\ \subset S$ be the corresponding
sets of variables: $S_A = \{X_i: i\in A\}$. Suppose we are given
joint probability distributions $P_A, P_B,\ldots$ for these sets of
variables. What are the conditions for these to be the marginal
distributions of a single probability distribution $P(x_1,\ldots,x_n)$?
This means that if, for example, $A=\{1,\ldots,r\}$, then 
\[ P_A(x_1,\ldots,x_r) = \sum_{x_{r+1},\ldots,x_n}P(x_1\ldots,x_n) 
\]
which we write as 
\[ P_A = \Sigma_{S \setminus A}(P).
\]
There are some obvious necessary conditions:
\begin{equation}\label{obvious}
 \Sigma_B(P_{A\cup B}) = \Sigma_C(P_{A\cup C}) \quad \text{if}
\quad A\cap B = A\cap C = \emptyset.
\end{equation}
In particular, $P_A$ is determined by $P_S$ if $A\subset S$. We may
therefore assume that in our given set of subsets, none is contained in
another. We will say that the subset distributions are \emph{equimarginal} if
they satisfy the conditions \eqref{obvious}. We ask what further conditions must be
satisfied.

The simplest non-trivial case --- which we discuss separately, for ease of
reading, even though it is contained in the general case which follows
--- is where $S$ is a set of three binary variables
and $A,B,C$ are the three two-element subsets, so that we are
considering three marginal two-variable distributions $P_{12}(x,y)$,
$P_{13}(x,z)$ and $P_{23}(y,z)$ where $x,y,z\in\{0,1\}$. Wigner  \cite{Wigner:ineq}
pointed out that these must satisfy
\be \label{Wigner}
P_{12}(x,y) \le P_{13}(x,z) + P_{23}(y,\ol{z})
\end{equation}
where $\ol{z}=1-z$ (but these inequalities are not satisfied by the predictions
of quantum mechanics for the measurements of the spin components of an electron
in three directions, where joint measurements in two different
directions are performed by measuring two electrons in a singlet state).
Pitowsky  \cite{Pitowsky:book} showed that the inequalities
\eqref{Wigner}, and the inequalities related to them by permuting
(1,2,3), are
sufficient for $P_{12}(x,y)$, $P_{13}(x,z)$ and $P_{23}(y,z)$ to be the
marginals of a single three-variable distribution $P(x,y,z)$.

To put these inequalities in a form which has a quantum analogue, we
regard $P_{12}(x,y)$ as a function of three variables $x,y,z$ which
is constant in $z$, and similarly for $P_{13}(x,z)$ and $P_{23}(y,z)$.
Then the functions $P_{12}, P_{13}, P_{23}$ are equimarginal if they satisfy three
equations like  
\[ 
P_{12}(x,y,z) + P_{12}(x,\ol{y},z) = P_{13}(x,y,z) +
P_{13}(x,y,\ol{z}) = P_1(x,y,z)
\]
where $P_1$ is constant in $y$ and $z$.

Now the observation of Wigner and Pitowsky can be expressed in terms of
three-variable functions as 

\begin{theorem}\label{3class}
Three equimarginal two-variable functions of three binary variables,
$P_{12}$, $P_{13}$ and $P_{23}$, are the two-variable marginals of a
three-variable probability distribution if and only if
\be\label{ineq3class}
0 \le \Delta(x,y,z) \le 1 \qquad \text{for all}\quad x,y,z\in\{0,1\}
\end{equation}
where
\[
\Delta = 1 - P_1 - P_2 - P_3 + P_{12} + P_{13} + P_{23}.
\]
\end{theorem}

\begin{proof} For $x\in\{0,1\}$, define $\sigma(x) = (-1)^x$, and write
\be 
\sigma_1(x,y,z) = \sigma(x), \quad \sigma_2(x,y,z) = \sigma(y), 
\quad \sigma_3(x,y,z) = \sigma(z).
\end{equation}
 Then any probability distribution 
$P$ on $\{0,1\}^3$ can be written
\be\label{sigmarep}
P = \tfrac{1}{8} + a\sigma_1 + b\sigma_2 + c\sigma_3 
+ d\sigma_1\sigma_2 + e\sigma_1\sigma_3 + f\sigma_2\sigma_3
+ g\sigma_1\sigma_2\sigma_3
\end{equation}
for some real constants $a,\ldots,g$. The marginals of $P$ are given by
\begin{align}\label{marginals}
P_{12} &= \quarter + 2a\sigma_1 + 2b\sigma_2 + 2d\sigma_1\sigma_2,\notag\\
P_{13} &= \quarter + 2a\sigma_1 + 2c\sigma_3 + 2e\sigma_1\sigma_3,\\
P_{23} &= \quarter + 2b\sigma_2 + 2c\sigma_3 + 2f\sigma_2\sigma_3\notag
\end{align}
and \upline
\[
P_1 = \half + 4a\sigma_1, \quad P_2 = \half + 4b\sigma_2, \quad P_3 =
\half + 4c\sigma_3.
\]
Hence 
\be
\Delta = \quarter + 2(d\sigma_1\sigma_2 + e\sigma_1\sigma_3 +
f\sigma_2\sigma_3),
\end{equation}
i.e. \upline
\be\label{Delta}
\Delta(x,y,z)= P(x,y,z) + P(\ol{x},\ol{y},\ol{z}).
\end{equation}
It follows that the inequality \eqref{ineq3class} is a necessary
condition for the existence of the probability distribution $P(x,y,z)$.

To prove that it is sufficient, note that the equimarginal condition 
forces the $P_{ij}$ to be of the form \eqref{marginals}.
We have to prove that there is a value of $g$ such that $P$ defined by
\eqref{sigmarep} is a positive function. Let
\[
Q = \tfrac{1}{8} + a\sigma_1 + b\sigma_2 + c\sigma_3 + d\sigma_1\sigma_2 +
e\sigma_1\sigma_3 + f\sigma_2\sigma_3;
\] 
then the conditions on $g$ are
\be 
- Q(x,y,z) \le g \le 1 - Q(x,y,z) 
\quad\quad\text{if }\;\sigma(x)\sigma(y)\sigma(z) = 1,\label{gineq1}
\end{equation}
and
\be
-1 + Q(x,y,z) \le g \le Q(x,y,z) 
\quad\quad\text{if }\;\sigma(x)\sigma(y)\sigma(z) = -1.\label{gineq2}
\end{equation}
But 
\[
Q = \quarter(P_{12} + P_{13} + P_{23} + \Delta) - \tfrac{1}{8}.
\]
Hence the condition $0\le\Delta\le 1$, together with $0\le P_{ij}\le 1$,
gives
\[ 
-\tfrac{1}{8} \le Q(x,y,z)\le \tfrac{7}{8}.
\]
It follows that every lower bound is less than every upper bound in
\eqref{gineq1} for different values of $(x,y,z)$; the same is true of 
\eqref{gineq2}; and every lower bound in \eqref{gineq2} is less than every
upper bound in \eqref{gineq1}.

Now suppose that $\sigma(x)\sigma(y)\sigma(z) = 1$ and
$\sigma(x\pr)\sigma(y\pr)\sigma(z\pr) = -1$. Then in the equations
\[
\sigma(x) = \pm\sigma(x\pr),\quad \sigma(y) = \pm\sigma(y\pr), \quad
\sigma(z) = \pm\sigma(z\pr)
\]
either one or three of the signs are negative. If all three are
negative, then
\begin{align*}
Q(x,y,z) + Q(x\pr,y\pr,z\pr) &= \quarter + 2\big(d\sigma(x) + e\sigma(x)\sigma(z) +
f\sigma(y)\sigma(z)\big)\\
&= \Delta(x,y,z).
\end{align*}
If just one sign is negative, say the first, then 
\begin{align*}
Q(x,y,z) + Q(x\pr,y\pr,z\pr) &= \quarter + 2\big(b\sigma(y) + c\sigma(z) + f\sigma(y)\sigma(z)\big)\\
&= P_{23}(y,z).
\end{align*}
In both cases we have
\be\label{Qineq1}
Q(x,y,z) + Q(x\pr,y\pr,z\pr) \ge 0
\end{equation}
so that every lower bound in \eqref{gineq1} is less than every upper
bound in \eqref{gineq2}. Thus there is a $g$ satisfying all of these
inequalities and giving the required probability distribution
$P(x,y,z)$.
\end{proof}

A classical probabilist would (probably) find it more natural to prove
necessity from the inclusion-exclusion principle, which gives $1 -
\Delta(x,y,z)$ as the probability that $X_1 = x$ or $X_2 = y$ or $X_3 = z$.
We have given our rather clumsier proof because it connects both with 
the proof of sufficiency and with the quantum problem.

We now move on to the general case of $n$ binary variables
$x_1,\ldots,x_n$. Let $N = \{1,\ldots,n\}$; for subsets
of $N$, we write $A\subset B$ to mean that $A$ is a proper
subset of $B$, writing $A\subseteq B$ when we want
to allow $A=B$; and $|A|$ denotes the number of elements of $A$.

We consider
probability distributions $P_A$ for subsets $A\subset N$, regarding
$P_A$ as a function of $(x_1,\ldots,x_n)$ which is constant in $x_i$ for
$i\notin A$. If $P(x_1,\ldots,x_n)$ is a probability distribution on all
$n$ variables, its marginal distributions $P_A$ can be written in terms
of operators $M_i$ on functions of $n$ binary variables defined by
\[ 
M_if(x_1,\ldots,x_n) = f(x_1,\ldots,x_n) +
f(x_1,\ldots,\ol{x_i},\ldots,x_n).
\] 
Then 
\[
P_A = M_{i_1}\ldots M_{i_r}P \qquad \text{where} \quad N \setminus A =
\{i_1,\ldots,i_r\}. 
\]
The distribution $P(x_1,\ldots,x_n)$ can be expanded as 
\begin{equation} \label{sigmaexpansion}
P = \sum_{A\subseteq N}c_A\sigma_A
\end{equation}
where the $c_A$ are real coefficients, with $c_\emptyset = 2^{-n}$, and 
\[
\sigma_A(x_1,\ldots,x_n) = \prod_{i\in A}\sigma(x_i), \qquad
\sigma_\emptyset = 1.
\]
Then the corresponding expansion of the marginal $P_A$ is 
\begin{equation}\label{marginal}
P_A = 2^{n-|A|}\sum_{B\subseteq A}c_B\sigma_B.
\end{equation}
This equation can be inverted to give $c_A\sigma_A$ in terms of the
marginals $P_A$:

\begin{equation}\label{sigmaterm}
c_A\sigma_A = \sum_{B\subseteq A}\frac{(-1)^{|A|-|B|}}{2^{n-|B|}}P_B.
\end{equation}

We can now state the generalisation of \thmref{3class} to
any number of variables:  

\begin{theorem}\label{nclass}  Let $P_A$ ($A\subset N$) be an equimarginal set of
probability distributions on subsets of the variables $x_1,\ldots,x_n$.
These are the marginals of a single distribution $P(x_1,\ldots,x_n)$ if
and only if for each
subset $A\subseteq  N$ with an odd number of elements,
\begin{equation}\label{genTony}
0\; \le \sum_{\substack{A\cup B = N\\\\B\subset N}}(-1)^{|A\cap B|}P_B(\x)
\;\le\; 1
\end{equation}
for all $\,\x\in\{0,1\}^n$.
\end{theorem}

\begin{proof}

To prove that the condition is necessary, suppose the distribution $P$
exists and let $A$ be a subset of $N$ with an odd number of elements.
Let $\x = (x_1,\ldots,x_n)$ and let $\x\pr$ be the sequence which
differs from $\x$ just in places belonging to $A$:
\[ 
x_i\pr = \begin{cases}\ol{x_i} \text{ if } i\in A\\
x_i \text{ if } i\notin A.\end{cases}
\]
Then
\[
0\le P(\x) + P(\x\pr ) \le 1.
\]
Expanding $P$ as in \eqref{sigmaexpansion}, we have
\[
P(\x) + P(\x\pr) = 2 \sum_{|A\cap B|\text{ even}}c_B\sigma_B(\x).
\]
Using \eqref{sigmaterm}, we can express this in terms of the probability
distributions $P_B$; the result is the sum in \eqref{genTony}. 
This can be verified
by using \eqref{marginal} to expand \eqref{genTony}:
\begin{equation}\label{expansion}
\sum_{N\setminus A\,\subseteq\, B\,\subset\, N}(-1)^{|A\cap B|}P_B 
  = \sum_{N\setminus A \,\subseteq \,B\, \subset\, N}(-1)^{|A\cap B|}2^{n-|B|}
     \sum_{D\subseteq B}c_D\sigma_D
\end{equation}
in which the coefficient of $c_D\sigma_D$ is
\begin{align*}  
  \sum_{\substack{B\supseteq D\\\\N\setminus A \subseteq B \subset N}}(-1)^{|A\cap B|}2^{n-|B|}
   &= \sum_{m=|A\cap D|}^{|A|-1}(-1)^m2^{|A| - m}
     \begin{pmatrix}|A| - |A\cap D|\\m-|A\cap D|\end{pmatrix}\\
   &\phantom{=} \qquad (\text{ writing } m = |A\cap B|)\\&\\
   &= (-1)^{|A\cap D|}2^{|A\setminus D|}\left\{\(1-\half\)^{|A\setminus D|} -
\(-\half\)^{|A\setminus D|}\right\}\\&\\
   &=(-1)^{|A\cap D|}\left\{ 1 - (-1)^{|A\setminus D|}\right\},
\end{align*}
so the right-hand side of \eqref{expansion} is 
\[
2\sum_{|A\setminus D| \text{ odd}}c_D\sigma_D\; =\; 2\sum_{|A\cap D| \text{ even}}
c_D\sigma_D 
\]
since $|A|$ is odd. Thus if the distribution $P$ exists, the inequality
\eqref{genTony} must be satisfied for each subset $A$ with an odd number
of elements.

To show that these inequalities are sufficient for the existence of the
distribution $P$, we first note, as in \thmref{3class}, that the 
equimarginality of the distributions $P_A$
gives us coefficients $c_B$ such that 
\[ 
P_A = \sum_{B\subseteq A}c_B\sigma_B.
\]
We have to prove that the stated conditions are sufficient to ensure
that there is a coefficient $c_N$ such that 
\[
P = \sum_{A\subset N} c_A\sigma_A + c_N\sigma_N
\]
satisfies $0\le P(\x) \le 1$ for all $\x\in\{0,1\}^n$. Writing 
\[ 
Q(\x) = \sum_{A\subset N} c_A\sigma_A(\x),
\]
we therefore need to be able to satisfy the inequalities
\begin{equation}\label{cNineq1} 
- Q(\x) \le c_N \le 1 - Q(\x)  \quad  \text{ whenever }\; \sigma_N(\x) = 1 
\end{equation}
and 
\begin{equation}\label{cNineq2}
-1 + Q(\x\pr) \le c_N \le Q(\x') \quad \text{whenever }\; \sigma_N(\x\pr) = -1. 
\end{equation}
Using \eqref{sigmaterm}, we can express $Q(\x)$ in terms of the
distributions $P_A(\x)$ as
\begin{align}
Q &= \sum_{A\subset N}\sum_{B\subseteq A} 
\frac{(-1)^{|A|-|B|}}{2^{n-|B|}}P_B \notag\\&\notag\\
  &= \sum_{B\subset N}\frac{P_B}{2^{n-|B|}}\sum_{B\subseteq A \subset N}
     (-1)^{|A|-|B|}\notag\\&\notag\\
  &= \sum_{B\subset N}\frac{P_B}{2^{n-|B|}}\sum_{m=|B|}^{n-1}
     (-1)^{m-|B|}\begin{pmatrix}n-|B|\\m-|B|\end{pmatrix}\notag\\&\notag\\
  &= \sum_{B\subset N}\frac{(-1)^{n-|B|-1}}{2^{n-|B|}}P_B.\label{Jason}
\end{align}

We will now show that the inequalities \eqref{genTony} imply 
\begin{equation}\label{Qineq} 
-\frac{1}{2^n} \le Q \le 1 - \frac{1}{2^n}.
\end{equation}
Indeed, summing these inequalities over all subsets $A$ with an odd
number of elements (of which there are $2^{n-1}$) gives
\[
0\;\le\; \sum_{B\subset N} d_B P_B(\x)\; \le\; 2^{n-1}
\]
where 
\begin{align*}
d_B &= \sum_{\substack{A\cup B = N\\\\|A|\text{ odd}}}(-1)^{|A\cap B|}\\
&\\
&= \sum_{r=0}^{|B|}\sum_{s\text{ odd}}(-1)^r\begin{matrix}\\\text{(number of $s$-element
subsets $A$ with $|A\cap B| = r$}\\\text{and $A\cup B = N$}\end{matrix}\\
&\\
&= \sum_{\substack{r=0\\\\n-|B|+r\text{ odd}}}^{|B|}(-1)^r
\begin{pmatrix}|B|\\r\end{pmatrix} \\
&\\
&=\begin{cases}1 \text{ if $B = \emptyset$ and $n$ is odd}\\
0 \text{ if $B = \emptyset$ and $n$ is even}\\
 (-1)^{n-|B|+1}2^{|B|-1}\;\text{ otherwise}\end{cases}
\end{align*}
since the sum of every other binomial coefficient in the $m$\/th row of
Pascal's triangle is $2^{m-1}$ if $m \ge 1$. Hence
\[ 0\; \le\; \frac{1}{2} + \sum_{B\subset
N}(-1)^{n-|B|+1}2^{|B|-1}P_B\; \le\; 2^{n-1}
\]
which, together with \eqref{Jason}, gives \eqref{Qineq}.
 
It follows from \eqref{Qineq} that if the inequalities \eqref{genTony} are satisfied, 
then every lower bound is less than every
upper bound in \eqref{cNineq1}, and therefore it is possible to satisfy all of these
inequalities with a single choice of $c_N$; and the same is true of
\eqref{cNineq2}. 

To be able to satisfy both sets of inequalities
simultaneously, we need 
\[
0 \le Q(\x) + Q(\x\pr) \le 2 \qquad \text{whenever } \sigma(\x) = 1
\text{ and } \sigma(\x\pr) = -1.
\]
If $\sigma(\x) = 1$ and $\sigma(\x\pr) = -1$, $\x$ and $\x\pr$ must
differ in an odd number of places. Let $A$ be the set of indices $i$
such that $x_i \neq x_i\pr$; then $\sigma_B(\x) = -\sigma_B(\x\pr)$ if
and only if $|A\cap B|$ is odd, so 
\[
Q(\x) + Q(\x\pr) = 2\sum_{|A\cap B| \text{ even}}c_B\sigma_B(\x),
\]
which, as we have already shown, is equal to the sum in \eqref{genTony}.
Hence if \eqref{genTony} is satisfied, then 
$Q(\x) + Q(\x\pr) \ge 0$, so no lower
bound in \eqref{cNineq1} is greater than any upper bound in
\eqref{cNineq2}; and $Q(\x) + Q(\x\pr) \le 2$, so 
no lower bound in \eqref{cNineq2} is
greater than any upper bound in \eqref{cNineq1}. It follows that it is possible to
find a suitable coefficient $c_N$, i.e. the conditions are sufficient
for the existence of a distribution $P$.
\end{proof}

The proof of this theorem suggests an alternative set of necessary and
sufficient conditions. Define the ``bit flip" operator $\kappa_i$ on
functions of $n$ binary variables $x_i\in\{0,1\}$ by 
\begin{equation}
(\kappa_i f)(x_1,\ldots,x_n) = f(x_1,\ldots,x_{i-1},\ol{x_i},x_{i+1},\ldots,x_n).
\end{equation}
and for any subset $A = \{i_1,\ldots,i_r\}$, let $\kappa_A =
\kappa_{i_1}\cdots\kappa_{i_r}$. Then

\begin{theorem}\label{classJason}  Let $P_A$ ($A\subset N$) be an equimarginal set of
probability distributions on subsets of the variables $x_1,\ldots,x_n$.
These are the marginals of a single distribution $P(x_1,\ldots,x_n)$ if
and only if, for all $\,\x\in\{0,1\}^n$,
\begin{equation}\label{genJason}
-\frac{1}{2^n} \le Q(\x) \le 1 - \frac{1}{2^n}
\end{equation}
and, for each odd subset $A\subset\{1,\ldots,n\}$,
\begin{equation}\label{genJason2}
0\le Q(\x)  + \kappa_A Q (\x) \le 2
\end{equation}
where 
\[
Q = \sum_{A\subset N}\frac{(-1)^{n-|A|-1}}{2^{n-|A|}}P_A.
\] 
\end{theorem}

\begin{proof} If the distribution $P(x_1,\ldots,x_n)$ exists, then we can expand it in terms of the functions $\sigma_A$ for subsets $A\subset N$ as in \eqref{sigmaexpansion}, and we have $Q(\x) = P(\x) - c_N\sigma_N(\x)$, as in 
\eqref{Jason}. The inequalities \eqref{cNineq1} and \eqref{cNineq2} follow, giving 
\[
0 \le Q(\x) + Q(\x\pr) \le 2 \qquad \text{whenever } \quad \sigma(\x) = 1
\text{ and } \sigma(\x\pr) = -1.
\]
This is equivalent to \eqref{genJason2}. Moreover, if $P$ exists then \thmref{nclass} holds and the inequalities \eqref{genJason} follow, as was shown in the proof of \thmref{nclass}.

Conversely, the stated inequalities on $Q$ guarantee that every left-hand side is less than every right-hand side in both \eqref{cNineq1} and \eqref{cNineq2}, and therefore there exists a coefficient $c_N$ such that $P = Q + c_N\sigma_N$ is a probability distribution. As in \thmref{nclass}, the equimarginal distributions $P_A$ can be expanded as 
\[
P_A(\x) = \sum_{B\subseteq A}c_B\sigma_B(\x)
\]
and then, by \eqref{Jason}, 
\[ 
Q(\x) = \sum_{B\subset N} c_B\sigma_B(\x)\quad\text{ where }\quad P_A(\x) = \sum_{B\subseteq A}c_B\sigma_B(\x)
\]
Hence the marginal distribution of $P = Q + c_N\sigma_N$ over the subset $A$ is 
\[
\Sigma_{N\setminus A}(P) = \Sigma_{N\setminus A}(Q) = \sum_{B\subseteq A}c_B\sigma_B = P_A,
\]
as required.
\end{proof}

\section{Quantum reduced states}

The general quantum problem concerns subsystems of a multipartite
system, with state space $\H=\H_1\ox\cdots\ox\H_n$ where $\H_1,\ldots,\H_n$
are the state spaces of the individual parts of the system. For each
subset $A\subset N = \{1,\ldots,n\}$, we denote the state space of the
corresponding subsystem by $\H_A = \bigotimes_{i\in A}\H_i$. Then the
problem is: Given a set of subsets $A,B,\ldots$ and states
$\rho_A,\rho_B,\ldots$ (density matrices on $\H_A,\H_B,\ldots$), does
there exist a state $\rho$ on $\H$ whose reduction to $\H_A$ is
$\rho_A$, i.e. 
\begin{equation}\label{reduced}
\rho_A = \tr_{\bar{A}}(\rho)\;?
\end{equation}
(Here $\bar{A}$ is the complement of $A$ in $\{1,\ldots,n\}$, and
tr$_{\bar{A}}$ denotes the trace over $\H_{\bar{A}}$ in the
decomposition $\H = \H_A \ox \H_{\bar{A}}$.) The
obvious compatibility conditions, corresponding to the classical
conditions \eqref{obvious}, are
\begin{equation}\label{qobvious}
\tr_B(\rho_{A\cup B}) = \tr_C(\rho_{A\cup C}) 
\quad \text{if} \quad A\cap B = A\cap C = \emptyset.
\end{equation}
As in the classical case, we will call a set of states \emph{equimarginal} if
they satisfy these conditions, and we can assume that none of the
subsets $A,B,\ldots$ is a subset of any other.

There is a further question in the quantum case: as well as asking whether there is
\emph{any} overall state with the given subsystem states as reduced
states, one can ask whether there is a pure state with this
property. This problem has a simplest case for which the classical and
mixed problems are trivial: if the given marginals are those of
all one-element subsets, then one can always construct the
classical probability distribution 
\[ f(x_1,\ldots,x_n) = f_1(x_1)\ldots f_n(x_n)
\]
with one-variable marginals $f_1,\ldots f_n$, and one can always
construct the quantum multipartite mixed state
\[ \rho = \rho_1\ox\cdots\ox\rho_n
\]
with one-party reduced states $\rho_1,\ldots,\rho_n$ (though not for
fermions: see the appendix). But it is not
always possible to find a pure state with these reductions. For a set of
qubits, necessary and sufficient conditions were found in
 \cite{polygon}:

\begin{theorem} Let $\rho_1,\ldots,\rho_n$ be a set of one-qubit density
matrices, and let $\lambda_i$ be the smaller eigenvalue of $\rho_i$.
Then there is an $n$-qubit pure state $|\Psi\>$ with one-qubit reduced
states $\rho_1,\ldots,\rho_n$ if and only
if $\lambda_1,\ldots,\lambda_n$ satisfy the polygon inequalities
\begin{equation}\label{polygon}
\lambda_i \le \sum_{j\ne i}\lambda_j\,.
\end{equation}
\end{theorem}

This result has been extended and generalised by a number of authors. Details are given
in the appendix.

Now let us consider the conditions for the existence of a mixed state
with given reduced states. The simplest case, as for the classical
problem, is a system of three qubits for which we are given two-qubit
reduced states $\rho_{12}, \rho_{13}, \rho_{23}$. The form in which we
have given the classical necessary and sufficient conditions can be 
immediately translated into quantum conditions by
replacing probability distributions by density matrices, and
inequalities between functions (holding for all values of the variables)
by inequalities between expectation values of operators, holding for all
states --- that is, positivity conditions on operators. We can prove
that this results in necessary conditions for the quantum problem, and
we conjecture that they are also sufficient. 

We will regard the reduced density matrix of a subsystem as an operator on the full
system by supposing that it acts as the identity on the remaining
factors of the full tensor product state space. That is, for three
qubits, we identify $\rho_{12}$ with $\rho_{12}\ox\mathbf{1}$, $\rho_2$
with $\mathbf{1}\ox\rho_2\ox\mathbf{1}$, etc. Then we have

\begin{theorem}\label{3quant} \emph{{\bf Quantum Bell-Wigner inequalities}} 
Suppose $\rho_{12}, \rho_{13}, \rho_{23}$
are the two-qubit reductions of a three-qubit mixed state. Then 
\[ 
0\le \<\Psi|\Delta|\Psi\> \le 1 
\]
for all normalised pure three-qubit states $|\Psi\>$, where
\[
\Delta = \mathbf{1} - \rho_1 - \rho_2 - \rho_3 + \rho_{12} + \rho_{13} +
\rho_{23}.
\]
\end{theorem}

\begin{proof}
This can be proved in a similar way to the classical version,
\thmref{3class}, with the help of the antiunitary ``universal NOT"
operator $\tau$ defined for one qubit by 
\[
\tau(a|0\> + b|1\>) = a^*|1\> - b^*|0\>.
\]
This operator satisfies $\tau^2 = -\1$ and 
anticommutes with all three Pauli operators
$\sigma_i$ ($i=1,2,3$). It is antiunitary, i.e.
\begin{equation}\label{antiuni}
\tau|\phi\> = |\ol{\phi}\>,\; \tau|\psi\> = |\ol{\psi}\> \; \impl 
\<\ol{\phi}|\ol{\psi}\> = \<\phi|\psi\>^*.
\end{equation}
We extend this to three-qubit states and define
\begin{equation}\label{tau}
\tau\(\sum_{\alpha\beta\gamma}c_{\alpha\beta\gamma}|\alpha\>|\beta\>|\gamma\>\)
 = (-1)^{\alpha + \beta + \gamma}c_{\alpha\beta\gamma}^*
|\ol{\alpha}\>|\ol{\beta}\>|\ol{\gamma}\>
\end{equation}
where $\alpha, \beta, \gamma \in \{0,1\}$ and $\ol{\alpha} = 1 -
\alpha$, etc.
This three-qubit operator is also antiunitary and squares to $-\1$,
which implies the ``universal NOT" property that it takes every pure
state to an orthogonal state. It anticommutes with the single-qubit Pauli
operators $\sigma_i\ox\operone\ox\operone$,
$\operone\ox\sigma_j\ox\operone$ and $\operone\ox\operone\ox\sigma_k$.

Any three-qubit mixed state can be written as 
\begin{align}\label{expandrho}
\rho &= \eighth\operone + a_i\sigma_i\ox\operone\ox\operone +
b_j\operone\ox\sigma_j\ox\operone + c_k\operone\ox\operone\ox\sigma_k\\
&\phantom{=} + d_{ij}\sigma_i\ox\sigma_j\ox\operone +
e_{ik}\sigma_i\ox\operone\ox\sigma_k +
f_{jk}\operone\ox\sigma_j\ox\sigma_k +
g_{ijk}\sigma_i\ox\sigma_j\ox\sigma_k \notag
\end{align}
(using the summation convention for
repeated indices), with real coefficients $a_i,\ldots,g_{ijk}$. 
The reduced states of $\rho$ are
\begin{align}\label{reductions}
\rho_{12} &= \quarter\operone + 2a_i\sigma_i\ox\operone + 
2b_j\operone\ox\sigma_j + 2d_{ij}\sigma_i\ox\sigma_j,\notag\\
\rho_{13} &= \quarter\operone + 2a_i\sigma_i\ox\operone + 
2c_k\operone\ox\sigma_k + 2e_{ik}\sigma_i\ox\sigma_k,\\
\rho_{23} &= \quarter\operone + 2b_j\sigma_j\ox\operone + 
2c_k\operone\ox\sigma_k + 2f_{jk}\sigma_j\ox\sigma_k\notag
\end{align}
and \upline
\[
\rho_1 = \half\operone + 4a_i\sigma_i, \quad 
\rho_2 = \half\operone + 4b_j\sigma_j, \quad 
\rho_3 = \half\operone + 4c_k\sigma_k.
\]
Hence 
\begin{align*}
\Delta &= \quarter\operone + 2(d_{ij}\sigma_i\ox\sigma_j\ox\operone + 
e_{ik}\sigma_i\ox\operone\ox\sigma_3 +
f_{jk}\operone\ox\sigma_j\ox\sigma_k)\\
&= \rho + \tau^{-1}\rho\tau
\end{align*}
since $\tau$ anticommutes with single-qubit Pauli operators. Thus
\begin{align}
\<\Psi|\Delta|\Psi\> &= \<\Psi|\rho|\Psi\> +
\<\ol{\Psi}|\rho|\ol{\Psi}\> 
\quad \text{where} \quad |\ol{\Psi}\> = \tau|\Psi\>\label{qDelta}\\
&\ge 0 \quad \text{ since $\rho$ is a positive operator.}\notag
\end{align}
Since $|\ol{\Psi}\>$ is orthogonal to $|\Psi\>$, \eqref{qDelta} also
gives
\[
\<\Psi|\Delta|\Psi\> \le \tr\rho = 1,
\]
establishing the theorem.
\end{proof}

We conjecture that the condition $0\le\Delta\le 1$ is also sufficient
for the existence of a three-qubit state with marginals $\rho_{12},
\rho_{13}, \rho_{23}$.

In the general multipartite case, the classical compatibility conditions of \thmref{nclass} also have quantum analogues, namely
\begin{equation}\label{genqTony}
0 \le \sum_{\substack{A\cup B = N\\\\B\subset N}}
(-1)^{|A\cap B|}\<\Psi|\rho_B|\Psi\> \le 1
\end{equation}
where $A\subseteq N$ is a subset with an odd number of elements. 
However, for $n > 3$ these are not even necessary conditions for compatibility (except for the case $A = N$, $n$ odd \cite{Bill:reduction}). The proof given above for $n=3$ fails because the universal-NOT operator $\tau$ is antilinear, not linear (which has the consequence that $\tau\otimes\1$ does not commute with $\1\otimes\sigma_i$). We illustrate this failure with a counter-example for $n=4$. In this case \eqref{genqTony} becomes 
\begin{equation}\label{qineq4}
0\le \<\Psi|\Delta_i|\Psi\> \le 1, \qquad i = 1,2,3,4
\end{equation}
where 
\[
\Delta_1 = \rho_1 - \rho_{12} - \rho_{13} - \rho_{14} + \rho_{123} + \rho_{124} + \rho_{134}
\]
and $\Delta_2,\Delta_3,\Delta_4$ are defined similarly. But consider
\[
\rho = |\Psi\>\<\Psi| \quad \text{ where } \quad  |\Psi\> = \tfrac{1}{\sqrt{2}}(|0000\> + |1100\>).
\]
We find that for this state
\[
\Delta_1 = \half(\1\otimes\1 - 2P_+)\otimes P_1\otimes P_1 + P_+\otimes P_0 \otimes P_0
\]
where $P_+$ is the two-qubit projector onto the maximally entangled state $\tfrac{1}{\sqrt{2}}\(|00\> + |11\>\)$, and $P_0$ and $P_1$ are the one-qubit projectors onto $|0\>$ and $|1\>$. Thus $\Delta_1$ has a negative eigenvalue $-\half$ with eigenvector $\tfrac{1}{\sqrt{2}}(|0011\> + |1111\>)$.

Since the classical inequalities are satisfied by all classical states, it is not surprising to find that the quantum analogues like \eqref{qineq4} are satisfied by separable states \cite{Bill:reduction}. Thus they constitute a set of separability criteria. These multipartite versions of the reduction criterion \cite{Hor2reduction,Cerf:reduction} have been investigated by Hall \cite{Bill:reduction}.

\linespace

\noindent\large\textbf{Acknowledgement}\quad\normalsize
 We are grateful to Sam Braunstein, whose
remark about the Wigner inequalities set us going in the right direction.

\appendix
\section{Appendix: Beyond Qubits}

Tripartite systems made up of state spaces with dimensions $d_i$ not all
equal to 2 have been studied by Higuchi ($d_1=d_2=d_3$) and Bravyi
($d_1=d_2=2, d_3=4$), who have found necessary conditions for a set of one-party
mixed states to be the reductions of a pure tripartite state. 
Their results are as follows:

\begin{theorem} \emph{(Higuchi \cite{Atsushi:3qutrit})} Three $3\times 3$ hermitian
matrices $\rho_a$ $(a = 1,2,3)$ with eigenvalues
$\lambda_1^{(a)}\le\lambda_2^{(a)}\le\lambda_3^{(a)} = 
1 - \lambda_1^{(a)} - \lambda_2^{(a)}$ are the reduced one-qutrit states
of a pure three-qutrit state if and only if
\begin{align*}
\alpha_a &\le \alpha_b + \alpha_c,\\
\beta_a &\le \alpha_b + \beta_c,\\
\gamma_a &\le \alpha_b + \beta_c,\\
\delta_a &\le \delta_b + \delta_c,\\
\epsilon_a &\le \delta_b + \epsilon_c,\\
\zeta_a &\le \delta_b + \zeta_c,\\
\text{and } \quad \zeta_a &\le \epsilon_b + \eta_c
\end{align*}
\begin{align*}
\text{where}\qquad \qquad \alpha_a = \lambda_1^{(a)} + \lambda_2^{(a)}, \quad
\beta_a = &\lambda_1^{(a)} + \lambda_3^{(a)}, \quad
\gamma_a = \lambda_2^{(a)} + \lambda_3^{(a)},\\
\delta_a = \lambda_1^{(a)} + 2\lambda_2^{(a)},\quad
\epsilon_a = 2\lambda_1^{(a)} + \lambda_2^{(a)},&\quad
\zeta_a = 2\lambda_2^{(a)} + \lambda_3^{(a)},\quad
\eta_a = 2\lambda_3^{(a)} + \lambda_2^{(a)}
\end{align*}
and $\{a,b,c\} = \{1,2,3\}$ in any order.
\end{theorem}

\begin{theorem} \emph{(Bravyi \cite{Bravyi})} Let $\rho_1$ and $\rho_2$ be two
$2\times 2$ density matrices with eigenvalues
$\lambda_a\le\mu_a=1-\lambda_a\; (a = 1,2)$, and let $\rho_3$ be
a $4\times 4$ density matrix with eigenvalues
$\lambda_3\le\mu_3\le\nu_3\le\xi_3 = 1 - \nu_1 -\nu_2 - \nu_3$. Then
$\rho_1, \rho_2$ and $\rho_3$ are the reduced states of a pure state in
$\C^2\ox\C^2\ox\C^4$ if and only if
\begin{align*}
\lambda_a &\ge \lambda_3 + \mu_3 \quad (a = 1,2),\\
\lambda_1 + \lambda_2 &\ge 2\lambda_3 + \mu_3 + \nu_3,\\
\text{and } \qquad |\lambda_1 - \lambda_2| &\le \min\{\nu_3 -
\lambda_3,\; \xi_3 - \mu_3\}.
\end{align*}
\end{theorem}

The general version of this inequality has been found by Han, Zhang and
Guo \cite{HanZhangGuo}, who, however, only proved that it is necessary:

\begin{theorem} \emph{(Han, Zhang and Guo \cite{HanZhangGuo})} Let
$\rho_1,\ldots ,\rho_n$ be the reduced one-particle density matrices of
a pure state of a system of $n$ particles each with an $m$-dimensional state space. Let
$\lambda_i^{(a)} (i=1,\ldots ,n)$ be the eigenvalues of $\rho_a$, with
$\lambda_1^{(a)}\le\cdots\le\lambda_n^{(a)}$. Then for each pair $(a,b)$ of
distinct particles and for each $p=1,2,\ldots m-1$,
\[
\sum_{i=1}^p\lambda_i^{(a)} \le \sum_{i=1}^p\lambda_i^{(b)} +
\sum_{\substack{c=1\\c\neq a,b}}^n\sum_{i=1}^{m-1}\lambda_i^{(c)}.
\]
\end{theorem}
These results can now be seen as special cases of a very general theorem due to Klyachko \cite{Klyachko}. A pure state is a special case of a mixed state with a given spectrum $(1,0,\ldots,0)$. One can consider a mixed state with any given spectrum and then, given a set of one-particle states, ask whether there is a mixed many-particle state with that spectrum which yields the given one-particle states. Klyachko has shown how to obtain sets of linear inequalities which give necessary and sufficient conditions on the one-particle spectra. For systems larger than four qubits, there are thousands of inequalities. Klyachko's methods belong to symplectic geometry, and are similar to the methods he used to solve the long-standing problem of Horn, who asked ``What are the possible spectra of a sum of hermitian matrices with given spectra?" Daftuar and Hayden have used these methods to find the possible spectra of a single reduced state $\rho_A$ obtained from a bipartite state $\rho_{AB}$; their paper \cite{DaftuarHayden} contains a very readable introduction to the relevant ideas from algebraic topology and symplectic geometry. There is a surprising connection to the representation theory of the symmetric group, which was also found by Christandl and Mitchison \cite{MatthiasGraeme}; roughly speaking, their result is that if the spectra of $\rho_{AB}, \rho_A$ and $\rho_B$ approximate the ratios of row lengths of Young diagrams $\lambda, \mu, \nu$, each with $N$ boxes, then the representation of the symmetric group $S_N$ labelled by $\lambda$ must occur in the tensor product of the representations labelled by $\mu$ and $\nu$.

Finally, we note that for fermions there is a non-trivial compatibility
problem for the one-particle reduced states of a mixed state. The
solution is as follows:

\begin{theorem} \emph{(Coleman \cite{Coleman:fermions})} An $m\times m$ density
matrix $\rho$ is the reduced one-party state of a system of $n$ fermions
if and only if each of its eigenvalues $\lambda$ satisfies
$0\le\lambda\le 1/n$.
\end{theorem}


\end{document}